\documentclass[12pt]{article}

\usepackage{latexsym}
\usepackage[dvips]{epsfig}
\usepackage[bf,footnotesize]{caption}   
\begin{document}
\setlength{\unitlength}{1mm}

\newcommand{\be}{\begin{equation}}
\newcommand{\ee}{\end{equation}}
\newcommand{\bea}{\begin{eqnarray*}}
\newcommand{\eea}{\end{eqnarray*}}
\newcommand{\bean}{\begin{eqnarray}}
\newcommand{\eean}{\end{eqnarray}}

\newcommand{\e}{{\mathrm e}}
\newcommand{\bbox}{\bar{\phantom{!}\Box\phantom{!}}}
\newcommand{\Tr}{\phantom{!}\mbox{\bf Tr}\phantom{!}}
\newcommand{\tr}{\phantom{!}\mbox{\bf tr}\phantom{!}}
\newcommand{\ra}{\rangle}
\newcommand{\la}{\langle}

\newcommand{\derv}[2]{ \frac{d #1}{d #2}}
\newcommand{\dervs}[2]{ \frac{d^2 #1}{d #2^2}}

\newcommand{\n}[1]{\label{#1}}
\newcommand{\eq}[1]{Eq.(\ref{#1})}
\newcommand{\ind}[1]{\mbox{\bf\tiny{#1}}}
\renewcommand\theequation{\thesection.\arabic{equation}}

\newcommand{\nn}{\nonumber \\ \nonumber \\}
\newcommand{\nl}{\\  \nonumber \\}
\newcommand{\pr}{\partial}
\renewcommand{\vec}[1]{\mbox{\boldmath$#1$}}


\title{ \hfill {\small Alberta-Thy-08-03 }  \vspace*{1cm} \\
{\bf Interaction of a brane with a moving bulk black hole} }
\author{
Valeri Frolov\footnote{e-mail: frolov@phys.ualberta.ca} ,
Martin Snajdr\footnote{e-mail: msnajdr@phys.ualberta.ca} , and
Dejan Stojkovi\'{c}\footnote{e-mail: dstojkov@phys.ualberta.ca}
}
\maketitle
\noindent
\centerline{ \em
Theoretical Physics Institute, Department of Physics,}
\centerline{ \em University of Alberta,  Edmonton, Canada T6G 2J1}
\bigskip

\maketitle
\noindent

\begin{abstract}

We study  the interaction of an $n$-dimensional topological defect
($n$-brane) described by the Nambu-Goto action with a
higher-dimensional Schwarzschild black hole  moving in the bulk
spacetime. We derive the general form of the perturbation equations
for an $n$-brane in the weak field approximation and solve them
analytically in the most interesting cases. We specially analyze
applications to brane world models. We calculate the induced geometry
on the brane generated by a moving black hole. From the point of view
of a brane observer, this geometry can be obtained by solving
$(n+1)$-dimensional Einstein's equations with a non-vanishing
right hand side. We calculate the effective stress-energy tensor
corresponding to this `shadow-matter'. We explicitly show that  there
exist regions on the brane where a brane observer sees an apparent
violation of energy conditions. We also study the deflection of light
propagating in the region of influence of this `shadow matter'.

\end{abstract}
\vspace{.3cm}


\section{Introduction}\label{s1} \setcounter{equation}0

Topological defects play an important role in modern physics.
Monopoles, strings, domain walls etc. are known as classical
solutions of field theory models which admit non-trivial topology.
They can arise in a wide class of phenomenological models including
GUT theories \cite{ViSh} and some extensions of the electroweak
standard model \cite{ssv}. The interaction of various topological defects
with black holes has been studied before. In \cite{FrolovBHS} and
\cite{FrolovBHDW} interaction of black holes with cosmic strings and
domain walls in $(3+1)$-dimensional universe was studied. Here we
generalize this consideration to a case where both a brane and a bulk
space in which the brane is moving may have an arbitrary number of
dimensions. We assume that the higher-dimensional brane can be
described by the Nambu-Goto action. We will study  motion of such
$(n+1)$-dimensional object in a background of $(n+k+1)$-dimensional
Schwarzschild black hole. We consider situations where the black hole
is far from the brane so that one can use  a weak field
approximation.

We will keep our analysis as general as possible, however we
will devote special attention to the so-called brane world model
in which the brane has three spatial dimensions while the number of
bulk dimensions is arbitrary. It was proposed  recently that the
whole universe could just be a three dimensional domain wall (a
brane) embedded in a higher dimensional space. In this model, all
the standard model particles are localized on the brane while gravity
can propagate everywhere. In particular, black holes being
gravitational solitons can propagate in a higher dimensional bulk
space. To make the consideration more concrete we use the so called
ADD model \cite{ADD} in which the gravitational field of the brane is
neglected and extra dimensions are flat. An important generic feature
of this model is that the fundamental quantum gravity mass scale
$M_*$ may be very low  (of order  TeV) and the size of
the extra spatial
dimensions may be much larger than the Planck length ($\sim
10^{-33}$cm). In the most interesting version of the ADD model there
exist two extra spatial dimensions with the size $L$ of the order of
0.1 mm,  which is still allowed by the experiments testing the Newton
law at small distances. The gravitational radius $R_0$ of a black
hole of mass $M$ in the spacetime with $k$ extra dimensions is defined
by the relation $G^{(4+k)}\, M\sim R_0^{k+1}$, where
$G^{(4+k)}=1/M_*^{(k+2)}$ is the $(4+k)$-dimensional Newton coupling
constant. The minimal mass of the black hole is determined by the
condition that its gravitational radius coincides with its Compton
length $\sim 1/M$. The mass of such an elementary black hole is $M_*$.
For $M_*\sim TeV$ one has $R_*\sim 10^{-17}$cm. When $M\gg M_*$ the
higher dimensional mini black holes can be described by the classical
solutions of vacuum Einstein's equations. We
assume that the size of a black hole $R_0$ is much smaller than the
characteristic size of extra dimensions, $L$,  and neglect the effects of
the black hole deformation connected with this size.

Black holes in the bulk space may exist as a result  the early stage
of the evolution of the universe, for example as a result of the
recoil during the quantum evaporation of primordial black holes
created on the brane \cite{FrSt}. The final state of such
evaporating mini black holes is not known. Similarly to the early
studied case of evaporation of the mini black holes in the $4$-dimensional theory
with the Planck  scale quantum gravity, two different final states are
possible. Either a stable remnant of mass of the order of $M_*$ is
formed or the evaporation is complete. We assume
that bulk black holes are either those stable remnants or black hole
with mass $M \gg M_*$, so that they are relatively long living.

We start with the general solution of an unperturbed flat $n$-brane
in $(n+k+1)$-dimensional  flat spacetime. We then introduce a moving bulk
black hole as a source of perturbation. Using coordinate
transformations we work in the reference frame in which the $n$-brane
is moving in the background spacetime of a static black hole. We
then derive the equations of motion of a perturbed brane in the weak
field approximation. Using the method of Green's functions, we give
the form of the general solution for an $n$-brane in arbitrary number
of spacetime dimensions. We specially analyze the most interesting
case of a $3$-brane in $4+1$ and $5+1$ dimensional spacetime where
we give the explicit solutions and discuss their properties. The
solutions for higher number of extra dimensions can easily be
generated from these solutions.
These solutions allow us to calculate the energy transfered from
the black hole to brane excitations in this process of interaction.

We also  calculate the induced
geometry on the brane generated by a moving black hole. As considered
by a brane observer this geometry can be obtained by solving  the
$(n+1)$-dimensional Einstein's equations with a non-vanishing
right hand side. We calculate the effective stress-energy tensor
corresponding to this `shadow-matter'.
We show that there exist regions where a brane observer sees an
apparent violation of energy conditions, namely, there exist
regions with negative energy density and negative pressure.
We also derive formulas for deflection of light propagating in the
induced spacetime metric on the brane.

\section{Black hole metric}\label{s2}
\setcounter{equation}0

The metric of a static multi-dimensional black hole in
$(K+1)$-dimensional spacetime is
\be\n{2.1}
dS^2=-F dT^2+ {dR^2\over F} +R^2\, d\Omega^2_{K-1}\, ,\hspace{1cm}
F=1-\left({R_0\over R}\right)^{K-2}\, ,
\ee
where $K$ is the number of spatial dimensions,
$R_0$ is the gravitational radius and $d\Omega^2_{K-1}$ is the line
element on a unit $(K-1)$-dimensional sphere.
For $K=3$ this is a usual $4$-dimensional Schwarzschild
metric.  We shall  use the metric (\ref{2.1}) in the {\em
isotropic coordinates}
\be\n{2.2}
dS^2= -F dT^2+ A^2\, \left[d\sigma^2 +\sigma^2\, d\Omega^2_{K-1} \right]\, ,
\ee
where
\be\n{2.3}
\ln\sigma =\int{dR\over R\sqrt{F(R)}}\, ,\hspace{1cm}
A={R\over \sigma}\, .
\ee

In the weak field approximation when $R_0/R\ll 1$ one gets
\be\n{2.4}
R\sim \sigma+ {R_0^{K-2}\over 2(K-2)\sigma^{K-3}}\, ,\hspace{1cm}
A\sim 1+{R_0^{K-2}\over 2(K-2)\sigma^{K-2}}\, ,
\ee
and the metric takes the following asymptotic form
\be\n{2.5}
dS^2=g_{AB}\, dX^A\, dX^B\, ,
\ee
\be\n{2.6}
g_{AB}=(1+\Psi)\eta_{AB} +(K-1)\Psi\, \delta_A^0\delta_B^0
=\eta_{AB}+\Psi\,h_{AB}\, ,
\ee
where
\be\n{2.7}
h_{AB}=\eta_{AB}+(K-1)\, \delta_A^0\delta_B^0 \, ,
\hspace{1cm}
\Psi={R_0^{K-2}\over (K-2)\sigma^{K-2}}\, .
\ee
 It should be emphasized that the metric
(\ref{2.5})-(\ref{2.6}) describes gravitational field of any
compact static distribution of matter \footnote{See for example
non-topological solitons in brane world models in \cite{NTS}.}
since we are considering only the leading  terms. The metric
(\ref{2.6}) is a perturbation over the background flat metric
\be\n{2.8}
dS_0^2=-dT^2 +  dL^2_{K}\, ,
\ee
where
\be\n{2.9}
dL^2_{K}= (dX^i)^2+(dY^m)^2= d\sigma^2+\sigma^2\, d\Omega^2_{K-1} \, ,
\ee
\be\n{2.10}
\sigma^2=(X^i)^2+(Y^m)^2\, .
\ee
We denote by $X^i$, $i=1,\ldots,n$ the `standard' $n$ Cartesian
coordinates, and by $Y^m$, $m=n+1,\ldots,n+k=K$ the Cartesian
coordinates in `extra-dimensions'. This
splitting will be convenient later when we
consider an $n$-dimensional brane in the spacetime of the
$(K+1)$-dimensional black hole.

\section{A flat brane in a spacetime with a fixed point}\label{s3}
\setcounter{equation}0

Our goal is to study the equations of motion of a brane  in the
gravitational field of a black hole.
We keep  the
number of brane spatial dimensions, $n$, and the number of
extra dimensions, $k$, arbitrary.

In the lowest order, when  the gravitational field of the black hole is neglected,
$g_{AB}=\eta_{AB}$.
$\eta_{AB}$ is the $(n+k+1)$-dimensional Minkowski metric, and the brane is flat. It is convenient to introduce
an orthonormal ennuple, $N_{\hat{B}}^A$,
\be\n{3.1}
g_{AB} \, N_{\hat{C}}^A N_{\hat{D}}^B=\eta_{\hat{C}\hat{D}}\,
\ee
so that its
first $n+1$ vectors, $N_{\hat{\mu}}^A$ ($\mu=0,\ldots,n$), are tangent
to the brane, while the other $k$ vectors,
$N_{\hat{m}}^A$ ($m=n+1,\ldots,n+k$),
 are  orthogonal to the brane.
A world-line $\Gamma$ representing a
position of uniformly moving black hole is described by the equation
\be\n{3.2}
X_{\Gamma}^A=(x_0^{\hat{B}}+U^{\hat{B}}\, T)\, N_{\hat{B}}^A\, .
\ee
Here $T$ is the proper time parameter along the world line, $U^A$ is
the $(n+k+1)$-velocity, $\eta_{\hat{A}\hat{B}}U^{\hat{A}} U^{\hat{B}}=-1$,
and $X_0^A \equiv x_0^{{\hat B}} N_{\hat{B}}^A$ are the coordinates of the black hole position at $T=0$.

Consider first the case of one extra dimension.  There are two
possibilities.

(1) The black hole crosses the brane. In this case we
use  the ambiguity $T\to T+$const to put $T=0$ at the
moment when the black hole meets the brane. Using the Poincare group
$P(n+1)$ of coordinate transformations preserving the position of the
brane in the bulk space one can always put
\be\n{3.3}
x_0^{\hat{B}}=0\, ,\hspace{1cm}
U^{\hat{B}}=\cosh \beta \delta^{\hat{B}}_{\hat{0}}+\sinh \beta
\delta^{\hat{B}}_{\hat{n+1}}\, .
\ee
That is a projection of the point representing the black hole onto
the brane surface is located at the origin of the brane coordinates,
at $x^{\hat{1}}=\ldots =x^{\hat{n}}=0$. The black hole crosses the
brane at the moment $x^{\hat{0}}=0$ of the brane time, and is moving
orthogonally to the brane surface. $\beta$ is a rapidity parameter
related to the velocity $v$ as $v=\tanh\beta$.

(2) The black hole never crosses the brane. This is a special case
when the velocity of the black hole relative to the brane vanishes. In this case by using
transformations from $P(n+1)$ one can put
\be\n{3.4}
x_0^{\hat{\mu}}=0\, ,\hspace{0.5cm}
x_0^{\hat{n+1}}=b\, ,\hspace{0.5cm}
U^{\hat{B}}=\delta^{\hat{B}}_{\hat{0}}\,  ,
\ee
where $b$ is the distance between the black hole and the brane.

Similarly, in the case with two or more extra dimensions two cases are
possible.

(1) The black hole crosses the brane. We choose $T$ so this
happens at the moment $T=0$. It means that $x_0^{\hat{m}}=0$. Using
the Poincare group $P(n+1)$ of coordinate transformations preserving
the position of the brane in the bulk space we put $x^{\hat{\mu}}=0$
and $U^{\hat{\mu}}=\cosh\beta \delta^{\hat{B}}_{\hat{0}}$.
We use  the group of rotations $O(k)$ which preserve the position of
the brane to put $U^{\hat{B}}=\sinh\beta
\delta^{\hat{B}}_{\hat{n+1}}$. Thus we have
\be\n{3.5}
x_0^{\hat{B}}=0\, ,\hspace{0.5cm}
U^{\hat{B}}=\cosh \beta \delta^{\hat{B}}_{\hat{0}}+\sinh \beta
\delta^{\hat{B}}_{\hat{n+1}}\, .
\ee

(2) The black hole never crosses the brane. There exists a minimal
distance, $b$, between the black hole and the brane, which we call the
{\em impact parameter}. As earlier we can put $x_0^{\hat{\mu}}=0$. We
also can choose the $N_{\hat{n+2}}^A$ to be directed from the brane
to the position of the black hole when it is at the minimal distance
from the brane. There still exists a group $O(k-1)$ of rotations
which preserves the position of the brane and the direction of
$N_{\hat{n+2}}^A$. We use this freedom to choose the vector
$N_{\hat{n+1}}^A$ to coincide with the direction of the black hole
velocity. For this choice we have
\be\n{3.6}
x_0^{\hat{B}}=b\, \delta^{\hat{B}}_{\hat{n+2}}\, ,\hspace{0.5cm}
U^{\hat{B}}=\cosh \beta \delta^{\hat{B}}_{\hat{0}}+\sinh \beta
\delta^{\hat{B}}_{\hat{n+1}}\, .
\ee

It is easy to see that the expression (\ref{3.6}) is in fact the most
general one. The relations (\ref{3.3}--\ref{3.5}) for the other cases
can be obtained from (\ref{3.6}) by either taking the limit $\beta=0$ or
putting $b=0$.

The gravitational potential $\Psi$ entering the expression
(\ref{2.6}) for the gravitational field of the black hole depends on
the interval $\sigma$ between the position of the black hole and a
point in a spacetime, calculated along $T$=const surface.
Let us calculate $\sigma$ for a point on the brane.
Denote by $V^{\hat{B}}$ a vector
\be\n{3.7}
V^{\hat{B}}=\sinh \beta \delta^{\hat{B}}_{\hat{0}}+\cosh \beta
\delta^{\hat{B}}_{\hat{n+1}}\, .
\ee
This vector is orthogonal to $U^{\hat{B}}$ and hence it is tangent to
$T$=const plane.
The brane time $x^{\hat{0}}$ corresponding to a given $T$ can be found
from the equations
\be\n{3.8}
T\, U^{\hat{B}}+\lambda\, V^{\hat{B}}=x^{\hat{0}}\,
\delta^{\hat{B}}_{\hat{0}}\, .
\ee
This equation for $\hat{B}=\hat{n+1}$ gives
\be\n{3.9}
\lambda=-\tanh\beta\, T\, ,
\ee
while for $\hat{B}=\hat{0}$ it gives
\be\n{3.10}
T=\cosh\beta\, x^{\hat{0}}\, ,\hspace{0.5cm}
\lambda= \sinh\beta\, x^{\hat{0}}\, .
\ee
Using this results we easily find that
\be\n{3.11}
\sigma^2=\rho^2+b^2+\sinh^2\beta\, (x^{\hat{0}})^2\, ,
\ee
where
$\rho^2=x^{\hat{i}}x_{\hat{i}}$.
From this expression for $\sigma$ it follows that the induced metric on the brane  will be
`spherically symmetric', that is, it possesses the group $O(n)$
of symmetry.

Let $X^{A}$ be Cartesian coordinates in the reference frame where the
black hole is at rest. Then the components of the ennuple
$N_{\hat{B}}^A$ in this frame are
\be\n{3.12}
N_{\hat{0}}^A=\cosh\beta\, \delta_0^A+\sinh\beta\, \delta_{n+1}^A\, ,
\hspace{1cm}
N_{\hat{i}}^A=\delta_i^A\, ,
\ee
\be\n{3.13}
N_{\hat{n+1}}^A=\sinh\beta\, \delta_0^A+\cosh\beta\, \delta_{n+1}^A\, ,
\hspace{1cm}
N_{\hat{m}}^A=\delta_m^A\, , \ m>n+1\,.
\ee
In the reference frame of the black hole the brane equation is
\be\n{3.14}
{\cal X}_0^A=x^{\hat{\mu}}\, N_{\hat{\mu}}^A + b\, N_{\hat{n+2}}^A\, .
\ee

\section{Brane perturbation equation of motion}\label{s4}
\setcounter{equation}0

Consider a brane ${\cal X}^A(x^{\hat{\mu}})$ moving in the spacetime
with a
metric $g_{AB}(X^C)$. The induced metric on the brane (which we
assume can be described by the Nambu-Goto action) is
\be\n{4.1}
\gamma_{\hat{\mu}\hat{\nu}}=g_{AB}({\cal X}^C)\, {\cal X}^A_{,\hat{\mu}}
\, {\cal X}^B_{,\hat{\nu}}\, .
\ee
The brane equation of motion is
\be\n{4.2}
\left(\sqrt{-\gamma}\, \gamma^{\hat{\mu}\hat{\nu}}\, \, {\cal X}^A_{,\hat{\mu}}\right)_{,\hat{\nu}}
+\sqrt{-\gamma}\, \gamma^{\hat{\mu}\hat{\nu}}\,{\Gamma^A}_{BC}\, {\cal
X}^B_{,\hat{\mu}}\,
{\cal X}^C_{,\hat{\nu}}=0\, .
\ee
When the brane is far from the black hole $g_{AB}$ has the form
(\ref{2.5}).

In the absence of gravity the unperturbed brane is described by the equation
(\ref{3.14}).
We write the solution for a perturbed brane in the form
\be\n{4.3}
{\cal X}^A={\cal X}_0^A+\chi^{\hat{m}}(x)\, N^A_{\hat{m}}
+\zeta^{\hat{\mu}}(x)\, N^A_{\hat{\mu}}\, .
\ee
Let us show first that by changing the coordinates on the brane one can
put $\zeta^{\hat{\mu}}=0$. Indeed, a change of Cartesian coordinates on
the brane $x^{\hat{\mu}}\to x^{\hat{\mu}}+\xi^{\hat{\mu}}(x)$
generates in (\ref{4.3}) an extra term
$\xi^{\hat{\mu}}(x)N^A_{\hat{\mu}}$. That is why, by using the
 diffeo-invariance of the brane equations and choosing
$\xi^{\hat{\mu}}=-\zeta^{\hat{\mu}}$ one can always take
\be\n{4.4}
{\cal X}^A={\cal X}_0^A+\chi^{\hat{m}}(x)\, N^A_{\hat{m}}\, .
\ee
In this
gauge $\chi^{\hat{m}}(x^{\hat{\nu}})$ are the `physical' degrees of freedom of
the brane. In a general case they describe both types of the brane
perturbations, free waves and brane deformations induced by an external
force. We focus our attention on the perturbations
induced by the brane motion in the weak gravitational field. We
restrict ourselves by considering only first order effects.

Using the relation
\be\n{4.5}
{\cal X}^A_{,\hat{\mu}}=N^A_{\hat{\mu}}+
\chi^{\hat{m}}_{,\hat{\mu}}\, N^A_{\hat{m}}\, ,
\ee
and keeping only the first order terms we obtain
\be\n{4.6}
\gamma_{\hat{\mu}\hat{\nu}}=\eta_{\hat{\mu}\hat{\nu}}+
\Psi \, h_{\hat{\mu}\hat{\nu}}\, ,\hspace{0.5cm}
h_{\hat{\mu}\hat{\nu}}=h_{AB}\, N^A_{\hat{\mu}}\, N^B_{\hat{\nu}}=
\eta_{\hat{\mu}\hat{\nu}}+
(K-1)\cosh^2\beta\, \delta_{\hat{\mu}}^0\delta_{\hat{\nu}}^0\, .
\ee
We also have
\be\n{4.7}
\sqrt{-\gamma}=1+{1\over 2}\, \left[ n+1-(K-1)\cosh^2\beta \right]\,
\Psi\, ,
\ee
\be\n{4.8}
\gamma^{\hat{\mu}\hat{\nu}}=\eta^{\hat{\mu}\hat{\nu}}-
\Psi\, h^{\hat{\mu}\hat{\nu}}\, ,\hspace{1cm}
h^{\hat{\mu}\hat{\nu}}=\eta^{\hat{\mu}\hat{\alpha}}\eta^{\hat{\nu}\beta}\,
h_{\hat{\alpha}\hat{\beta}}\, .
\ee

In the chosen coordinate system, the Christoffel symbols
${\Gamma^A}_{BC}$ are first order quantities
\be\n{4.9}
{\Gamma^A}_{BC}=\eta^{AD}\, \Gamma_{D BC}\, ,\hspace{1cm}
\Gamma_{D BC}={1\over 2}\, \left( \Psi_{,B}\, h_{CD} +\Psi_{,C}\,
h_{BD}-\Psi_{,D}\, h_{BC} \right)\, .
\ee

By multiplying (\ref{4.2}) by $N_{\hat{m} A}$ we obtain the
following equations
\be\n{4.10}
\Box^{(n+1)}\, \chi_{\hat{m}}=f_{\hat{m}}\, ,
\ee
\be\n{4.11}
f_{\hat{m}}={1\over 2}\Psi_{,\hat{m}}\, h +{1\over 2}
(K-1)\sinh(2\beta)\Psi_{,\hat{0}}\,
\delta_m^{n+1}\, ,\hspace{0.5cm}
h=\eta^{\hat{\mu}\hat{\nu}}h_{\hat{\mu}\hat{\nu}}=n+1-(K-1)\cosh^2\beta\, .
\ee
Here
 $\Psi_{,\hat{\mu}}=N^{A}_{\hat{\mu}}\Psi_{,A}$,
$\Psi_{,\hat{m}}=N^{A}_{\hat{m}}\Psi_{,A}$, and
the $(n+1)$-dimensional flat `box'-operator is defined as
\be\n{4.12}
\Box^{(n+1)}=\eta^{\hat{\mu}\hat{\nu}}\partial_{\hat{\mu}}
\partial_{\hat{\nu}}\, .
\ee
It is  easy to check that the other equations obtained by
multiplication of (\ref{4.2}) by $N_{\hat{\mu} A}$ are
trivially satisfied.

To calculate $f_{\hat{m}}$ we note that the function $\Psi$ which
enters this expression depends on $\sigma^2$, i.e. $\Psi (\sigma^2)$, so that
\be\n{4.13}
\Psi_{,\hat{0}}=\Psi'\, \sigma^2_{,\hat{0}}\, ,\hspace{1cm}
\Psi_{,\hat{m}}=\Psi'\, \sigma^2_{,\hat{m}}\, ,
\ee
where
\be\n{4.14}
\Psi'= \frac{\partial \Psi}{\partial \sigma^2} = -{R_0^{K-2}\over 2\sigma^K}\, ,
\ee
\be\n{4.15}
\sigma^2=\rho^2+(\sinh^2\beta ) t^2+b^2\, .
\ee
We denoted $t=x^{\hat{0}}$, $\rho^2=x^{\hat{i}}x_{\hat{i}}$.
We also have
\be\n{4.16}
\sigma^2_{,\hat{0}}=2t\sinh^2 \beta\, ,\hspace{0.5cm}
\sigma^2_{,\hat{n+1}}=2t\sinh\beta\, \cosh\beta\, , \hspace{0.5cm}
\sigma^2_{,\hat{n+2}}=2b\, .
\ee
The other terms $\sigma^2_{,\hat{n+p}}$ with $p>2$ vanish.

The induced metric on the unperturbed brane in the
spherical coordinates is
\be\n{4.17}
ds_0^2=\eta_{\hat{\mu}\hat{\nu}}dx^{\hat{\mu}}dx^{\hat{\nu}}=-dt^2+d\rho^2+\rho^2\,
d\Omega_{n-2}^2\, .
\ee
Because of the symmetry of the problem, the `force' terms $f_{\hat{m}}$ on
the right hand side of (\ref{4.10}) are functions of $t$ and $\rho$.
Thus the induced perturbations of the brane are `spherically symmetric'
and can be written in the form
\be\n{4.18}
\chi_{\hat{m}}={P_{m}(t,\rho)\over \rho^{(n-1)/2}}\, .
\ee
By substituting this expression into (\ref{4.10}) one obtains
\be\n{4.19}
\left[ -\partial_t^2+\partial_{\rho}^2 -
{(n-1)(n-3)\over 4 \rho^2} \right] P_{m}(t,\rho)= F_{m}(t,\rho)\, ,
\ee
where
\be\n{4.20}
f_{\hat{m}}={F_{m}(t,\rho)\over \rho^{(n-1)/2}}\, \, .
\ee

From this equation we see that that the cases of a string ($n=1$)
and a three-brane ($n=3$) are particularly easy to
study. Solutions for the general case are discussed in the appendix
\ref{appA}.

\section{Solutions of the brane perturbation equations}\label{s5}
\setcounter{equation}0

\subsection{Generators of solutions}

In the chosen coordinate system and imposed gauge, the `force'
$f_{\hat{m}}$ which enters the right hand side of (\ref{4.11}) has
only two non-vanishing components, $f_{-}=f_{\hat{n+1}}$ and
$f_{+}=f_{\hat{n+2}}$. Since we consider only solutions which are
induced by the `force' acting on the brane, we shall also have only two
non-trivial functions, $\chi_{-}=\chi_{\hat{n+1}}$ and
$\chi_{+}=\chi_{\hat{n+2}}$ which describe the brane excitations.
We write the equation (\ref{4.11}) as
\be\n{5.1}
\Box^{(n+1)}\chi^{(n,k)}_{\pm} = f^{(n,k)}_{\pm}\ .
\ee
We included an upper index $(n,k)$ to make clear the dependence of the
functions on  the number of  spatial dimensions, $n$, of the brane,
and  on the number of extra dimensions, $k$. Simple calculations give
\be\n{5.2}
f^{(n,k)}_{\pm} = A^{(n,k)}_{\pm}\, \tilde{f}^{(n,k)}_{\pm}\, ,
\ee
where
\be\n{5.3}
\tilde{f}^{(n,k)}_{-}=\frac{t}{\sigma^{n+k}}\,
,
\hspace{1cm}
\tilde{f}^{(n,k)}_{+}=\frac{1}{\sigma^{n+k}}\,
,
\ee
\be\n{5.5}
A^{(n,k)}_{-}=-\frac{1}{4}R_0^{n+k-2}\left[2-k+(n+k-1)\sinh^2\beta\right]\sinh(2\beta)\
,
\ee
\be\n{5.6}
A^{(n,k)}_{+}=-\frac{1}{2}R_0^{n+k-2}\left[n+1-(n+k-1)\cosh^2\beta\right]b\,
.
\ee
If we write
\be\n{5.7}
\chi^{(n,k)}_{\pm} = A^{(n,k)}_{\pm}\, \tilde{\chi}^{(n,k)}_{\pm}\, ,
\ee
then
\be\n{5.8}
\Box^{(n+1)}\tilde{\chi}^{(n,k)}_{\pm} = \tilde{f}^{(n,k)}_{\pm}\ .
\ee
Let us note now that
\be\n{5.9}
\tilde{f}^{(n,k+2)}_{\pm} = -\frac{2}{n+k}\
\frac{\partial}{\partial(b^2)}\tilde{f}^{(n,k)}_{\pm}\ ,
\ee
and therefore
\be\n{5.10}
\tilde{\chi}^{(n,k+2)}_{\pm} = -\frac{2}{n+k}\
\frac{\partial}{\partial(b^2)}\tilde{\chi}^{(n,k)}_{\pm}\ .
\ee
This relation shows that one can generate  solutions for an
arbitrary $k>2$ if the solutions
for $k=1,2$ are known.
It should be noted that for $k=1$ there is only one transverse
excitation of the brane, so that a solution
$\chi^{(n,1)}_{+}$ does not have a direct physical meaning.
Only its derivatives corresponding to higher values of $k$ are physical.
We call the functions $\chi^{(n,1)}_{\pm}$ and $\chi^{(n,2)}_{\pm}$
{\em generating solutions}.

\subsection{Generating solutions for $n=3$ brane}

As an important example we consider now a special case when the brane
has three spatial dimensions. This case is interesting for brane
world models. The generating solutions for this case can be written
as follows (see appendix~\ref{appA})
\be\n{5.11}
\tilde{\chi}^{(k)}_{\pm} = \frac{-1}{2\rho}\int_{-\infty}^{t}dt'\int_{0}^{\infty}d\rho'\
\rho'
\tilde{f}^{(k)}_{\pm}(\vartheta(\lambda_+)-\vartheta(\lambda_-))\,
, \ee
where
\be\n{5.12}
\lambda_{\pm}= (t-t')^2-(\rho \mp \rho')^2)\, .
\ee
In order to make notations more compact we omitted index $n=3$ in the
superscript.

\begin{figure}[h!]
\begin{center}
\epsfig{file=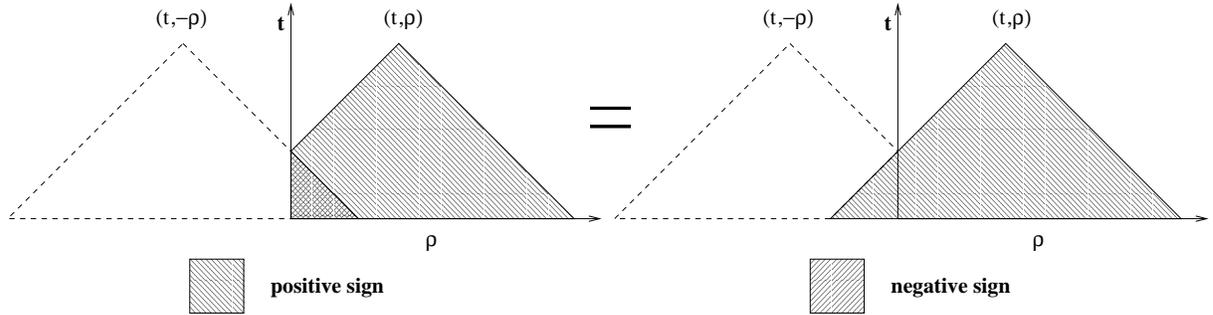, width=\textwidth}
\caption{It is equivalent to integrate over the two regions on the left
as it is to integrate over the one region on the right.}
\label{domain}
\end{center}
\end{figure}

Since the $\tilde{f}^{(k)}_{\pm}$ are even functions of $\rho$ the
integral over the two `mirror' regions can be rewritten as an
integral over one region characterized only by the
$\vartheta(\lambda_+)$ without the restriction of $\rho$ being
greater than zero (see figure~\ref{domain}).

After calculating the integrals we obtain
\be\n{5.13}
\tilde{\chi}^{(1)}_{-}=
\frac{1}{\cosh^2\beta}
\left[(t+\rho)\, S_+^{(1)} - (t-\rho)\, S_-^{(1)} \right]\ ,
\ee
\be\n{5.14}
\tilde{\chi}^{(1)}_{+}=S_+^{(1)} -S_-^{(1)},
\ee
where
\be\n{5.15}
S_{\pm}^{(1)}=
{1\over 4\rho R_{\pm}}\left[
\arctan{\left(\frac{t\sinh^2\beta \mp \rho}{R_{\pm}}\right)}+{\pi\over
2}\right]\, ,
\ee
and
\be\n{5.16}
R^2_\pm=(t\pm\rho)^2\sinh^2\beta+b^2\cosh^2\beta\ .
\ee
We also have
\be\n{5.17}
\tilde{\chi}^{(2)}_{-}=\frac{1}{\cosh^2\beta}
\left[
(t+\rho)\, S^{(2)}_+ -(t-\rho)\, S^{(2)}_-\right]\ ,
\ee
\be\n{5.18}
\tilde{\chi}^{(2)}_{+}=S^{(2)}_+ -S^{(2)}_-\, ,
\ee
where
\be\n{5.19}
S_{\pm}^{(2)}=
{1\over 6\rho R^2_{\pm}}\left(
\frac{t\sinh^2\beta\mp\rho}{\sqrt{\rho^2+t^2\sinh^2\beta+b^2}}
+\cosh\beta \right)\, .
\ee

\begin{figure}[h!]
\center
\begin{tabular}{cc}
\epsfig{file=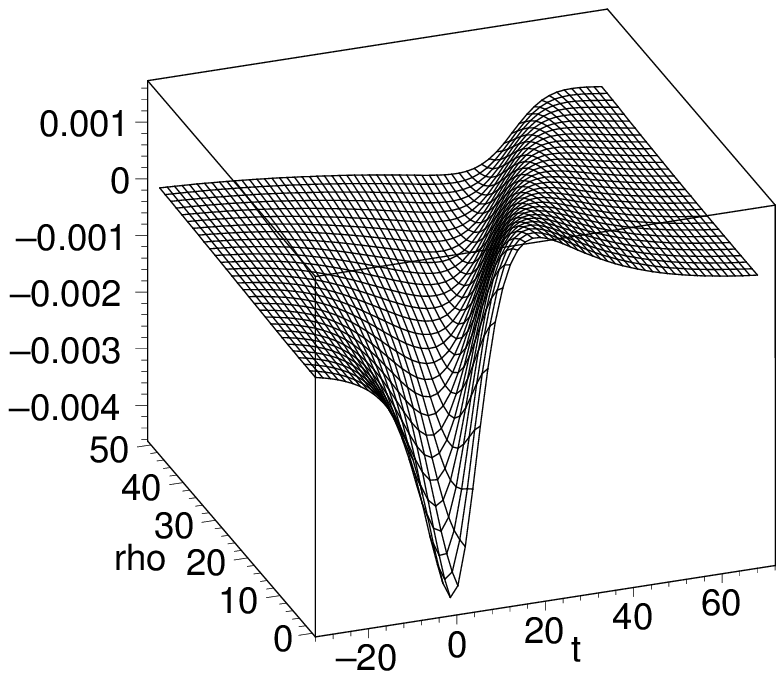} &
\epsfig{file=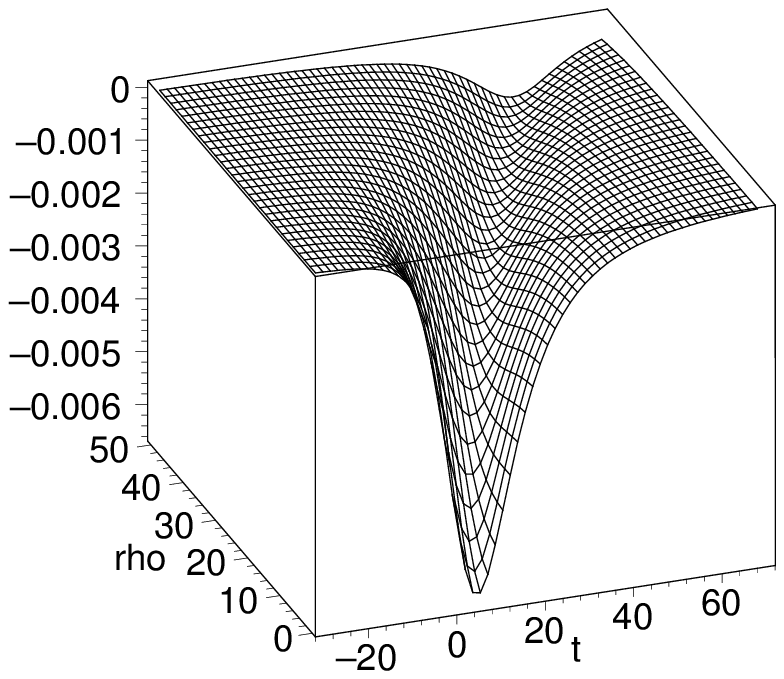}  \\
\bf{a) $\chi_-$} & \bf{b) $\chi_+$} \\
\end{tabular}
\caption{Plots of $\chi_-$ and $\chi_+$ for $\beta=1$, $R_0=1$, $b=10$, $k=2$.}
\label{solplot}
\end{figure}

Let us illustrate the motion of the brane in the case $n=3$, and
$k=2$. The parametric equations for the brane in isotropic
coordinates are
\be\n{5.20}
X^0 =
t\cosh\beta - 2 R_0^3\, \sinh^4\beta\, \cosh\beta\, \tilde{\chi}^{(2)}_{-}\ ,
\ee
\be\n{5.21}
X^1 = x^1\ ,\hspace{0.5cm}
X^2 = x^2\ ,\hspace{0.5cm}
X^3 = x^3\ ,
\ee
\be\n{5.22}
X^4 =
t\sinh\beta - 2 R_0^3\, \sinh^3\beta\, \cosh^2\beta\,
\tilde{\chi}^{(2)}_{-}\ ,
\ee
\be\n{5.23}
X^5 =
b + 2b \, R_0^3\, \sinh^2\beta\, \tilde{\chi}^{(2)}_{+}\ .
\ee
Figure~\ref{solplot} shows plots of $\chi_{\pm}$ for
a particular choice of parameters. Both plots depict a disturbance
of the brane developed around ``time'' $t=0$ which travels
at a speed of light outward from the
point of the brane closest to the black hole.

\subsection{Energy loss}

As a result of the black hole action, the brane transforms from its
initial state without excitations to a new excited  state. The energy
gained by the brane in this process is equal to the loss of the
kinetic energy of the black hole. We calculate now this energy loss.

\begin{figure}[h!]
\center
\begin{tabular}{cc}
\epsfig{file=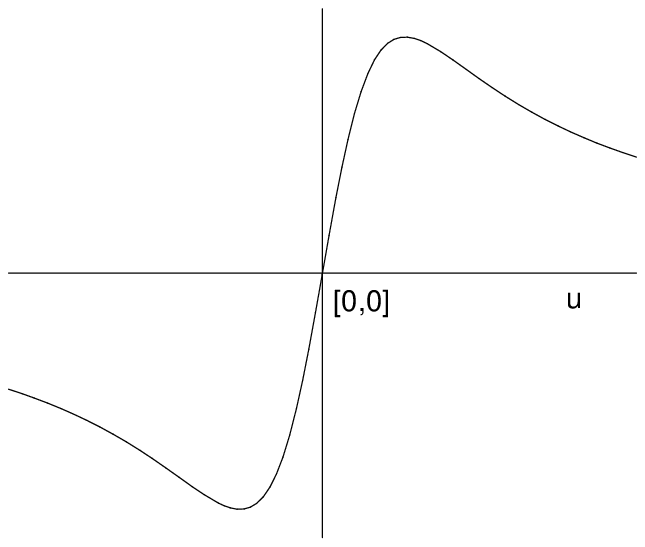} &
\hspace{1cm}
\epsfig{file=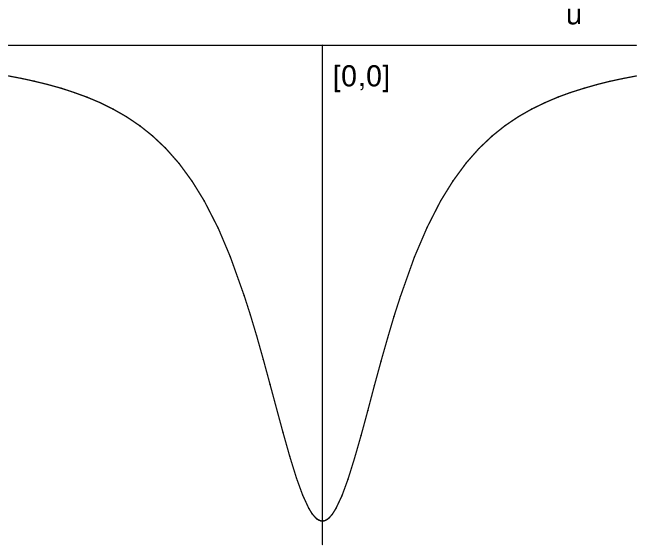}  \\
\bf{a) $\Phi_-(u)$} & \bf{b) $\Phi_+(u)$} \\
\end{tabular}
\caption{Asymptotic shape of the fields $\Phi_-$ and $\Phi_+$.}
\label{asympt_chi}
\end{figure}

In the limit $t\to\infty$, \ $u=t-\rho=$ const, the brane excitation
amplitudes $\chi_{\pm}$ take the following form
\be\n{e.1}
\chi_{\pm}={\Phi_{\pm}(u)\over \rho}\, ,
\ee
\be\n{e.2}
\Phi_- = \frac{2}{3}\,  R_0^3\sinh^3\beta\, \frac{u}{u^2\sinh^2\beta
+b^2\cosh^2\beta}
\ ,
\ee
\be\n{e.3}
\Phi_+ = -\frac{2}{3}\, R_0^3\,b\cosh\beta\sinh^2\beta\, \frac{1}{u^2\sinh^2\beta
+b^2\cosh^2\beta}
\ .
\ee
Figure \ref{asympt_chi} shows the typical shape of the functions $\Phi_{\pm}$
around $u=0$.

In the asymptotic regime, when the gravitational field of the black
hole can be neglected, the induced metric (\ref{4.1}) is
\be
\gamma_{\hat{\mu}\hat{\nu}}=\eta_{\hat{\mu}\hat{\nu}}+
\chi^{\hat{m}}_{, \hat{\mu}}\, {\chi_{\hat{m}}}_{, \hat{\nu}} \, .
\ee
The Nambu-Goto action is
\be
I=\mu\, \int \, \sqrt{-\gamma}\, d^n x =I_0 +I_2\, ,
\ee
where $\mu$ is the tension of the brane, $I_0$ is constant and
\be
I_2=-{1\over 2}\mu\, \sum_{\hat{m}=n+1}^{n+k}\,\int \,(\nabla \chi^{\hat{m}})^2 \, .
\ee
In our case the asymptotic excitations are described by two massless
scalar fields with the Lagrangian density
\be\n{e.4}
{\cal L}_{\pm} =
-\frac{1}{2}\mu\, \partial_\mu\chi_{\pm}\partial^\mu\chi_{\pm}\, .
\ee
The energy flux ${\cal E}_{\pm}$ calculated at the future null infinity
is
\be\n{e.5}
{\cal E}_{\pm}=4\pi \, \mu\, \int_{-\infty}^{\infty}\, du\,
({\Phi_{\pm}}_{,u})^2\,
.
\ee
Simple calculations give
\be\n{e.6}
{\cal E}_{-}={\cal E}_{+}=\frac{4 \,\pi^2\, \mu\, \sinh^5\beta}{9\cosh^3\beta}\,
\frac{R_0^6}{b^3}\, .
\ee
Thus the total energy lost by the black hole and gained by the brane
is
\be\n{e.7}
\Delta E={\cal E}_{-}+{\cal E}_{+}=\frac{8 \,\pi^2\, \mu\,
\sinh^5\beta}{9\cosh^3\beta}\, \frac{R_0^6}{b^3}\, .
\ee

Since extra dimensions are compactified, the black hole will be passing near
the brane again and again. Because of the friction force connected
with the energy loss, the black hole will be slowing down
until finally stops with respect to the brane.

\section{`Shadow matter' effect}\label{s6}
\setcounter{equation}0

The metric induced on the brane by a black hole moving in the bulk
space is
\be\n{6.1}
ds^2=-\left[ 1+[1-(k+2)\cosh^2\beta]\, \Psi  \right] dt^2
+(1+\Psi)\left[ d\rho^2+\rho^2\, d\Omega_{2}^2 \right]\, ,
\ee
with
\be
\Psi={R_0^{k+1}\over (k+1)\sigma^{k+1}}\, \hspace{.5cm} {\rm and }
 \hspace{.5cm} \sigma^2 = \rho^2 + (\sinh^2 \beta ) t^2 + b^2
 \, .
\ee

Particles and light on the brane are moving along geodesics in this
metric. If an observer on the brane does not know about the existence of
extra dimension and uses the standard 4-dimensional Einstein's equations
he would arrive to the conclusion that there exists some distribution
of matter on the brane responsible for this gravitational
field. Since this matter is not connected with any usual 4-dimensional physical
fields and particles  we call it  a `shadow matter'. We discuss now the
properties of this matter.

The Einstein tensor $G_{\hat{\mu}\hat{\nu}}=R_{\hat{\mu}\hat{\nu}}-
{1/2}\gamma_{\hat{\mu}\hat{\nu}}R$ for
the metric (\ref{6.1}) takes the form ( in spherical coordinates $(t,\rho, \theta, \phi)$)
\be\n{6.2}
G_{\hat{0}\hat{0}}={R_0^{k+1}\over \sigma^{k+5}}\left[
3t^2\sinh^2\beta-\rho^2k+3b^2\right]\, ,
\ee
\be\n{6.3}
G_{\hat{1}\hat{1}}={R_0^{k+1}\over \sigma^{k+5}}
\left[ k t^2 \sinh^2\beta +(\rho^2+b^2)((3+k)\cosh^2\beta-3)\right]\ ,
\ee
\be\n{6.4}
G_{\hat{0}\hat{1}}=-{t\rho R_0^{k+1}(3+k)\sinh^2\beta\over
\sigma^{k+5}}\, ,
\ee
\be\n{6.5}
G_{\hat{2}\hat{2}}={G_{\hat{3}\hat{3}}\over \sin^{2}\theta }=
{R_0^{k+1}\rho^2\over 2\sigma^{k+5}}
\left[k[2t^2 \sinh^2\beta+\rho^2(2-(3+k)\cosh^2\beta)]+2b^2((3+k)\cosh^2\beta-3)
\right]\, .
\ee

\begin{figure}[h!]
\center
\begin{tabular}{cc}
\epsfig{file=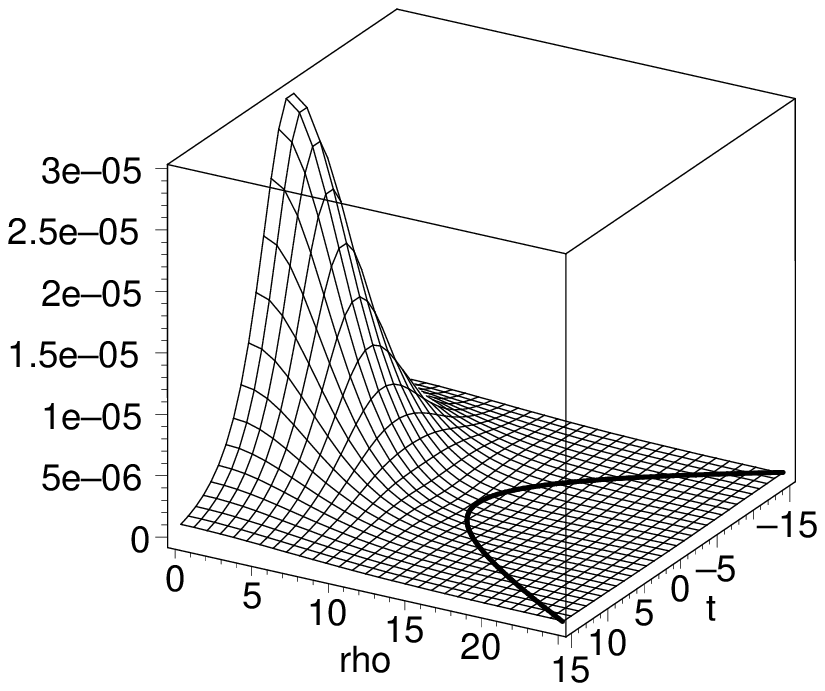} &
\epsfig{file=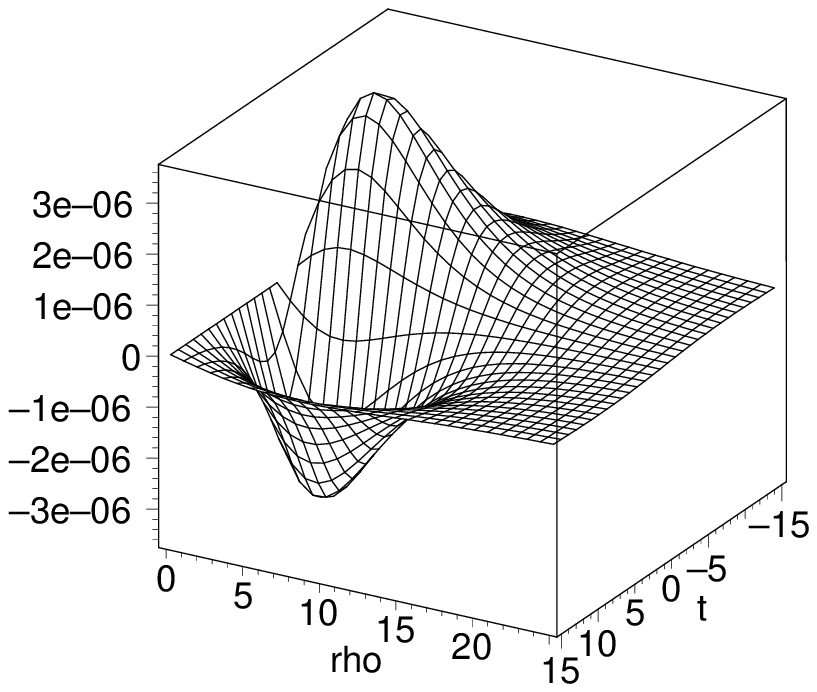}  \\
\bf{a) $G_{tt}$} & \bf{b) $G_{t\rho}$} \vspace{0.5cm} \\
\epsfig{file=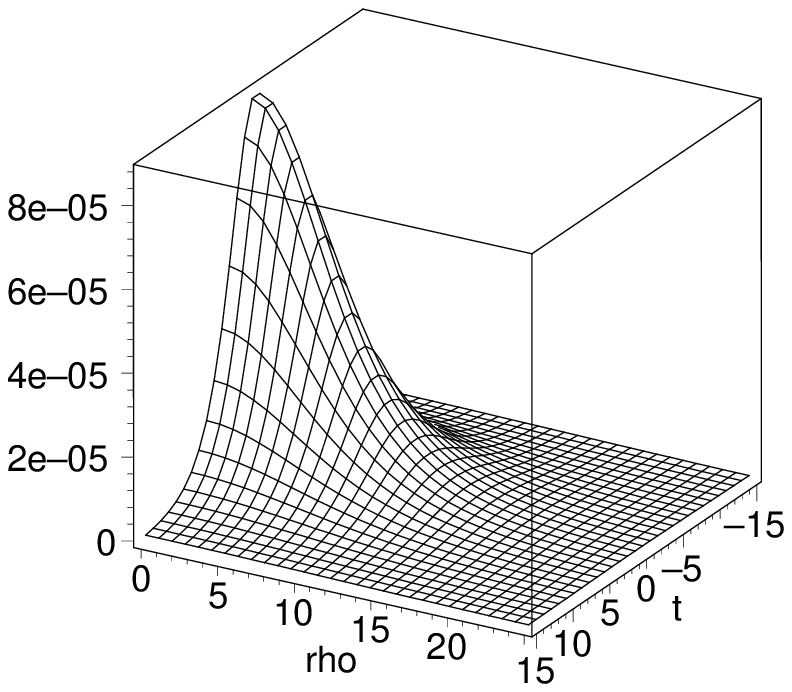} &
\epsfig{file=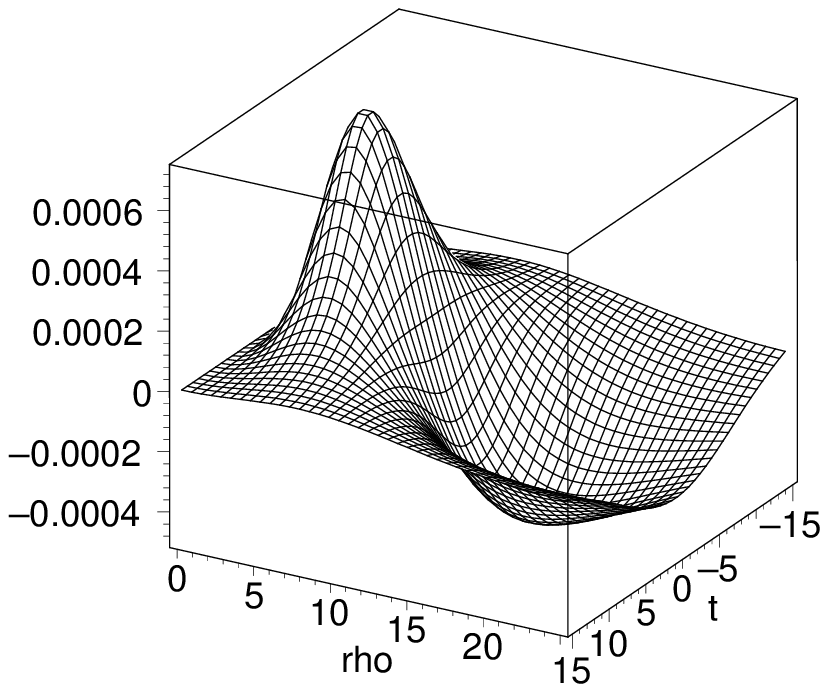}  \\
\bf{c) $G_{\rho\rho}$} & \bf{d) $G_{\theta\theta}$} \\
\end{tabular}
\caption{Plots of the components of the Einstein tensor of the
induced geometry on the brane. The parameters are  $R_0=1$, $\beta=1$, $b=10$, $k=2$}
\label{einstein}
\end{figure}

Figure~\ref{einstein} shows plots of the various non-zero components
of the Einstein tensor $G_{\mu\nu}$ which is proportional to the
stress-energy tensor $T_{\mu\nu}$ measured by observers living on the
brane. The thick line on figure~\ref{einstein}a marks the border
between the positive and negative energy density.

In the simplest case when  the bulk black hole is not moving,
$\beta=0$, these expressions simplify and non-vanishing components of
the Einstein tensor are (at $t=0$)
\be\n{6.6}
G_{\hat{0}\hat{0}}={R_0^{k+1}\over \sigma^{k+5}}\left[
-\rho^2 k+3b^2\right]\, ,
\ee
\be\n{6.7}
G_{\hat{1}\hat{1}}={R_0^{k+1}\over \sigma^{k+5}}
k(\rho^2+b^2)\, ,
\ee
\be\n{6.8}
G_{\hat{2}\hat{2}}={G_{\hat{3}\hat{3}}\over \sin^{2}\theta }=
{R_0^{k+1}\rho^2 k \over 2\sigma^{3+k}}
\left[ -\rho^2(k+1)+2b^2\right]\, .
\ee

Suppose a brane observer uses the Einstein's equations to describe the
gravitational field on the brane. In this case he would come to a
conclusion that there exists some form of matter for which
\be\n{6.9}
T_{\hat{\mu}\hat{\nu}}={1\over 8\pi G^{(4)}}\,
G_{\hat{\mu}\hat{\nu}}\, ,
\ee
where $G^{(4)}$ is a 4-dimensional Newtonian coupling constant.
For a static black hole this spherically symmetric distribution of
matter is of the form
\be\n{6.10}
T_{\hat{\mu}}^{\hat{\nu}}=\mbox{diag}(-\varepsilon,
p_\rho,p_{ \perp},p_{\perp})\, .
\ee

Since the total number
of spatial dimensions is greater than three, i.e.  $k>0$, the energy
density $\varepsilon$ is positive at the center and changes its sign
and becomes negative at $\rho>\rho_b=b\sqrt{3/k}$. The radial pressure
$p_\rho$ is always positive, while the transverse pressures
$p_{\perp}$ are positive at the center and
become negative at $\rho>b\sqrt{2/(k+1)}$.

If there is a distribution of physical matter on the brane, it will
give an additional contribution to the metric (\ref{6.1}). In the
weak field approximation this contribution will be additive.

Let us estimate the total positive mass $m_b$ of the `shadow matter'
inside the sphere of the radius $\rho_b$. To define this mass one can
use the relation
\be\n{6.11}
m(r)=4\pi\int_0^{r} \, dr\, r^2\, T_{\hat{0}\hat{0}} \, .
\ee
The radius coordinate $r$ is related to the isotropic coordinate as
\be\n{6.12}
(1+\Psi)\, \rho^2=r^2\, .
\ee
Since we are considering only first order terms we can write
\be\n{6.125}
m (\rho)={1\over 2 G^{(4)}} \, \int_0^{\rho} \, d\rho\, \rho^2\,
 G_{\hat{0}\hat{0}}
= \frac{1}{2 G^{(4)}}
 \frac{R_0 \rho^3}{(\rho^2+b^2)^{\frac{k+3}{2}}}\,
 \ee
In particular, we have
\be\n{6.13}
m_b= m(\rho_b) = \alpha(k)\, {R_0\over G^{(4)}}\, \left(R_0\over b \right)^k\, ,
\ee
where
\be\n{6.14}
\alpha(k)= {3\sqrt{3}\over 2}\, {k^{k/2}\over (k+3)^{(k+3)/2}}\, .
\ee
It is convenient to rewrite (\ref{6.13}) as
\be\n{6.15}
m_b=\alpha(k)\,  m_*\, \left({R_0\over R_*} \right)^{k+1}\,
\left({R_*\over b} \right)^{k}\, .
\ee
Here $R_*=1/M_*$ is the gravitational radius of the fundamental mini black
hole, and $m_*=R_*/G^{(4)}$. For $M_*\sim $TeV one has $m_*\sim
10^{11}$g! Thus, for $b\sim $TeV, we have $100 000$ tons
of ``shadow matter" concentrated
in the region of the size TeV$^{-1}$. However, this feature is
visible only for test particles whose wavelength is of order
TeV$^{-1}$.

The mass $m_b$ is surrounded by the negative mass distribution
$\varepsilon$. For infinite size of extra dimensions, at far
distances it exactly cancels the mass $m_b$, so that  the total mass
as measured at infinity vanishes. It happens because the induced
gravitational field potential falls down at infinity as it is
required by $(3+k)$-dimensional Newton's law (i.e. $\sim
G^{(4+k)}M/r^{1+k}$), while the standard $3$-dimensional Newtonian
gravitational potential of mass $M$ at far distance is $G^{(4)}M/r$.
For a finite size $L$ of extra dimensions this cancellation is not
complete. The gravitational field of the bulk mass $M$ as measured by
the brane observer at $r\gg L$ is $\sim G^{(4+k)}M/(L^k\,r)=
G^{(4)}M/r$. In other words, the bulk masses at $r\gg L$ contribute
to the gravitational field on the brane similarly to the matter on
the brane.

One can easily check that for  $\rho > \rho_b$ the `shadow
matter' distribution violates the weak energy condition \footnote{ In
fact the dominant and strong energy conditions are
also violated.}: $\varepsilon \geq 0$ and $ \varepsilon + p_i \geq
0$, $(i =1,2,3)$.  This violation is apparent only for an observer
located on the brane. The complete system (brane $+$ bulk) does not
violate any of the energy conditions. Similar effects can be present in
many different frameworks (see for example \cite{vec}). Our
situation is distinguishable by the fact that the source of apparent
violation of energy conditions is a simple distribution of bulk
matter.

The violation of the
weak energy condition in particular implies that  an $n$-dimensional
pencil of initially parallel null rays propagating on the brane and
passing through the region with $\rho > \rho_b$ will be defocused,
that is the  $n$-dimensional area of its cross-section will increase. If on the
other hand one consider the $(n+k)$-dimensional  pencil of initially
parallel null rays propagating in the bulk, the area of its $(n+k)$-dimensional
cross-section will decrease. There is no contradiction between these
two results, since the Weyl tensor of the bulk gravitational field
does not vanish. As a result a shear is generated, and the
expansion of the beam in the direction of the brane is compensated by
the contraction of the beam in the bulk dimensions.

\section{Deflection of light}
\setcounter{equation}0

The `shadow-matter' can affect the propagation of test particles
and light on the brane. We consider now the deflection of
 a light ray passing in the region of influence of the
`shadow-matter'. For simplicity we assume that the bulk black hole
velocity vanishes.

Since the propagation of light is invariant
under conformal transformation of the metric, we divide the metric
(\ref{6.1}) by $1+\Psi$ and keep the leading order terms
\be\n{7.1}
ds^2=-\left( 1+A \Psi  \right) dt^2
+ d\rho^2+\rho^2\,d\theta^2 + \rho^2 \sin^2\theta d\phi^2 \, ,
\ee
where $A = -(k+2)$.
Since the metric is spherically symmetric the light trajectory lies in
a plane and we can set $\theta =\pi/2$. The eiconal equation which
describes the motion of light in some background metric is
\be\n{7.2}
g^{\mu \nu} \frac{\partial S}{\partial x^\mu} \frac{\partial S}{\partial
x^\nu}=
-(1-A\psi) \left( \frac{\partial S}{\partial t}
\right)^2 + \left( \frac{\partial S}{\partial \rho}
\right)^2  + \frac{1}{\rho^2} \left( \frac{\partial S}{\partial \phi}
\right)^2=
0\, .
\ee
Here $S$ is the Hamilton-Jacobi action for massless particle (often
called eiconal).

The action $S$ can be written as
\be \n{7.3}
S = -\omega t + m \phi + S_\rho (\rho) \, ,
\ee
where $\omega$ is the energy and $m$ is the angular momentum.
Substituting (\ref{7.3}) into (\ref{7.2}) we get
\be\n{7.4}
S_\rho = \int \sqrt{(1- A\Psi)\, \omega^2 - \frac{m^2}{\rho^2}}  \, d\rho\ .
\ee
When the spacetime is flat the light ray is a straight line. The
corresponding $S_\rho^{(0)}$ is given by  the integral (\ref{7.4})
with $A=0$. Keeping  only the first order terms we can write
\be \n{7.5}
S_{\rho}=S_\rho^{(0)} + \Delta S_\rho\, ,
\ee
where
\be \n{7.6}
\Delta S_\rho =  \int
\frac{A \omega^2\, \Psi  }{ 2 \sqrt{ \omega^2 - \frac{m^2}{\rho^2} } }
 \, d\rho\ .
\ee
After substituting the value for $A$ and
\be \n{7.7}
\Psi = \frac{R_0^{k+1}}{(k+1)(\rho^2 + b^2)^{\frac{k+1}{2}}}
\ee
we obtain
\be\n{7.75}
\Delta S_\rho = - \int_R^{b_0}
\frac{(k+2) \omega R_0^{k+1} \rho d\rho }{ (k+1)
\sqrt{ (\rho^2+b^2)^{k+1} ( \rho^2 - \frac{m^2}{\omega^2} )} }\ .
\ee

Note that the impact parameter of the light ray in terms of integrals of
motion is just $b_0 = \frac{m}{\omega}$.
Light propagates from some large distance $\rho =
R$ to the point $\rho = b_0$ nearest to the center of the
projected `shadow matter' and then back to the distance $R$.
Therefore the integration goes from $R$ to $b_0$ and the
result should be multiplied by 2.
For a ray coming from infinity  $R\rightarrow \infty$.
After integration we get
\be\n{7.8}
\Delta S_\rho = \frac{(k+2)\sqrt{\pi}\omega \Gamma (\frac{k}{2})}{2
(k+1)\Gamma
(\frac{k+1}{2})} \frac{R_0^{k+1}}{(b_0^2+b^2)^{\frac{k}{2}}}\ .
\ee
From eq. (\ref{7.3}) we obtain the total deflection angle
\be\n{7.9}
\Delta \phi = -\frac{\partial}{\partial m} ( \Delta S_\rho )\ .
\ee
The calculation gives
\be\n{7.10}
\Delta \phi^{(k)} = \frac{(k+2)\, \sqrt{\pi}\,  k \, \Gamma (k/2)}{2 \,
(k+1)\, \Gamma((k+1)/2)}\, \,
\frac{R_0^{k+1}\,  b_0}{(b_0^2+b^2)^{\frac{k+2}{2}}}\ .
\ee
In particular, for the case of one and two extra dimensions one has
\be
\Delta \phi^{(1)} = \frac{3}{4} \pi R_0^2
\frac{b_0}{(b^2+b_0^2)^{3/2}}\, ,
\hspace{1cm}
\Delta \phi^{(2)} = \frac{8}{3} R_0^3
\frac{b_0}{(b^2+b_0^2)^{2}}\, .
\ee
For comparison,  the standard $(3+1)$-dimensional case ($k=0$, $b=0$)
is
\be
\Delta \phi = \frac{2 R_0}{b_0}
\ee
The different functional dependence of the deflection angle on the
impact parameter of light $b_0$ gives us opportunity to
distinguish between the real matter on the brane and the bulk
`shadow matter' as a cause of deflection.

\section{Conclusions}

We studied the interaction of an $n$-dimensional brane
described by the Nambu-Goto action with a
higher-dimensional Schwarzschild black hole  moving in the
($n+k+1$)-dimensional spacetime. The $n=1$ case corresponds to a cosmic string,
$n=2$ to a domain wall, while $n=3$ can be interpreted as the
observable universe in the context of brane world models.
We derived the general form of the perturbation equations
for an $n$-brane in the background of a $(n+k)$-dimensional
black hole in the weak field approximation.

For odd number of spatial brane dimensions by using convenient
mathematical transformations we derived the closed form solution of
the D'Alambert Equation with a spherically symmetric
source (see appendix~\ref{appA}). We applied this result to the most interesting
case of a $3$-brane in a spacetime with extra dimensions where
we obtained a general solution.

We calculated the induced geometry on the brane generated by a moving
black hole. As considered by a brane observer this geometry can be
obtained by solving $(n+1)$-dimensional Einstein's equations with a
non-vanishing right hand side. We calculated the effective
stress-energy tensor corresponding to this `shadow-matter'. We showed
that there exist regions where a brane observer sees an apparent
violation of energy conditions. The `shadow-matter' also affects the
propagation of test particles and light on the brane. We demonstrated
this by deriving results for deflection of light propagating in the
induced spacetime metric on the brane. As expected, results are
quite different from the $(3+1)$-dimensional results. It would be
interesting to study eventual  observational tests which would
indicate that the `shadow-matter' and thus the extra dimensions
influence the physics of our $(3+1)$-dimensional world.

One of the possible interesting application of the `shadow matter'
effect might be the following. If there exists a diluted
non-relativistic gas of stable elementary mini black holes in the
extra dimensions, its gravitational action on the brane would be
similar to the ``observable" dark matter, provided the
$(3+k)$-dimensional density of this gas is $n_k$ where
\be
\epsilon_{DM}\sim M_*\, n_k \, L^k\, .
\ee
Here $\epsilon_{DM}$
is the mass density of the dark matter. For  $\epsilon_{DM}\approx
10^{-29}$g/cm${}^3$ one has
\be
n_k \sim 10^{-6+2k} 1/cm^{3+k}\, .
\ee
Since the average distance between the bulk black holes,
$n_k^{-1/(3+k)}$, is much larger than the black hole radius
$R_*$, with a very high accuracy one can consider such gas to be very
diluted and neglect the gravitational interaction between the black
holes. A distinguishing property of this model is that there are no
physical particles on the brane responsible for the dark matter.

\bigskip

\vspace{12pt} {\bf Acknowledgments}:\ \  This work was partly
supported  by  the Natural Sciences and Engineering Research
Council of Canada. V.F. and D.S. are grateful to the Killam Trust
for its financial support.

\bigskip

\newpage

\appendix
\section{Spherically-Symmetric Solutions of D'Alambert Equation in
\label{appA}
Higher Dimensional Spacetimes}
\setcounter{equation}0

A solution of the $(n+1)$-dimensional D'Alambert equation
\be\n{a.1}
\Box^{(n+1)} p = f
\ee
can be written by using a retarded Green's function
\be\n{a.2}
\Box^{(n+1)} G^{ret}_{n+1}(x,x')=- \delta^{(n+1)}(x-x')\, .
\ee
We assume that in the infinite past there were no incoming waves.
The solution $p(x)$ is then completely generated by the external `force'
$f(x)$,
\be
p(x)=-\int G^{ret}_{n+1}(x,x')\, f(x')\, dx'\, .
\ee

The Green's function for odd $n=2\nu+1$, \ $\nu\ge 1$, can be written
as \cite{Galtsov}

\be\n{a.3}
G^{ret}_{n+1}(x,x')= {(-1)^{\nu -1}\vartheta(t-t')\over
(2\pi)^{\nu} } \
 \left[ {d^{\nu-1}\over
(R dR)^{\nu-1}}\, \delta((t-t')^2-R^2)\right]\, ,
\ee
where $R=|{\bf x}-{\bf x}'|$.
For our convenience, we rewrite it in a slightly different form

\be\n{a.35}
G^{ret}_{n+1}(x,x')= {\vartheta(t-t')\over 2\pi^\nu } \
\left[ {d^{\nu-1}\over
d\alpha^{\nu-1}}\, \delta(\lambda+\alpha)\right]_{\alpha=0}\, .
\ee
Here
\be\n{a.5}
\lambda =(t-t')^2-|{\bf x}-{\bf x}'|^2\ ,
\ee
\be\n{a.6}
|{\bf x}-{\bf x}'|^2 =\rho^2+\rho'^2-2\rho\rho'z\, ,\hspace{1cm}z= \cos\theta\, ,
\ee
and $\theta$ is the angle between $n$-dimensional vectors ${\bf x}$
and ${\bf x}'$.

We are interested  in spherically symmetric solutions of D'Alambert
equation. Integrating over the angular variables we get
\be\n{a.7}
\hat{G}(t,\rho;t',\rho')=\Omega_{n-2}\,
\int_{-1}^1\, dz\, (1-z^2)^{(n-3)/2} G^{ret}_{n+1}(t,\rho;t',\rho';z)\, ,
\ee
where $\Omega_k=2\pi^{\frac{k+1}{2}}/\Gamma(\frac{k+1}{2})$ is
a volume of a
$k$-dimensional unit sphere $S^k$. Here we made explicit that
$G^{ret}_{n+1}$ depends on the angle variables only through the
parameter $z$.

Let us denote
\be F=\rho^{\nu}\, f\, ,\hspace{1cm}P=\rho^{\nu}\, p\ .
\ee
Then in the absence of incoming waves the spherically
symmetric solution of the equation (\ref{a.1}) is
\be\n{a.8}
P(t,\rho)=- \int dt'\, d\rho'\,  {\cal G}(t,\rho;t',\rho')\,
F(t',\rho')\, ,
\ee
where
\be {\cal G}(t,\rho;t',\rho')=\left(
{\rho' \rho} \right)^{\nu} \, \hat{G}(t,\rho;t',\rho')\ .
\ee

Let us show that  for odd $n$ the representation (\ref{a.3})
allows one to obtain the Green's function for this reduced
equation. The reduced Green's function ${\cal G}$ can be written
as
\be\n{a.9} {\cal G}(t,\rho;t',\rho')=\vartheta(t-t')
{ \left( {\rho' \rho} \right)^{\nu} \over \Gamma(\nu )}\, B\, , \ee
where
\be\n{a.10} B=\,\left[ {d^{\nu-1}\over d\alpha^{\nu-1}}\,
\int_{-1}^1\, dz\, (1-z^2)^{\nu-1}
\delta(b(z+z_*))\right]_{\alpha=0}\, , \ee \be\n{a.11}
\lambda_0=(t-t')^2-\rho^2-\rho'^2\, ,\hspace{0.5cm}
b=2\rho\rho'\, , \hspace{0.5cm} z_*={\lambda_0+\alpha\over b}\,.
\ee
By calculating the integral in (\ref{a.10}) one gets \be\n{a.12}
B=B_+ - B_-\, , \ee \be\n{a.13} B_{\pm}={1\over b}\,\left[
{d^{\nu-1}\over d\alpha^{\nu-1}}\,\left\{ (1-z_*^2)^{\nu-1}\,
\vartheta(z_*\pm 1) \right\}\,\right]_{\alpha=0}\, . \ee
Now let us note that if any of the derivative over $\alpha$ is
acting on the $\vartheta$-function, the result vanishes because
of the remaining factor $1-z_*^2$. Using Rodrigues' formula for
Legendre polynomials (\cite{AS}, relation 22.11.5)

\be\n{a.14} P_n(x)={(-1)^n\over 2^n\, n!}\, {d^n\over dx^n}\left[
(1-x^2)^n\right]\, , \ee
one obtains
\be\n{a.141}
B_{\pm}=\vartheta(\lambda_{\pm})\, {(-2)^{\nu-1}\over b^{\nu}}\,
(\nu-1)!\, P_{\nu-1}(\lambda_0/b)\, , \ee where \be\n{a.15}
\lambda_{\pm}= (t-t')^2-(\rho \mp \rho')^2)\, . \ee
Combining the above expressions we obtain the following
representation for the reduced Green's function
\be\n{a.16} {\cal
G}(t,\rho;t',\rho')= C_{\nu}\, P_{\nu-1} \left(
\frac{\lambda_0}{b} \right) \left[ \vartheta(\lambda_+)-
\vartheta(\lambda_-)\right] \ ,
\ee
where
\be C_{\nu}=\frac{(-1)^{\nu-1}(\nu -1) !}{2 \Gamma (\nu)}\, . \ee
For $\nu =1$, which corresponds to a $n=3$ brane we have
\be
p(t, \rho) = -\frac{1}{2} \int dt' d\rho'\frac{\rho'}{\rho} f(t', \rho')
\left[ \vartheta(\lambda_+)-\vartheta(\lambda_-)\right]\, .
\ee

A similar procedure can be applied for the case of even $n$,
i.e., odd number of spacetime dimensions. The expressions are more
complicated and  will not be discussed here.

\end{document}